\documentstyle[11pt]{article}
\input psfig
\long\def\comment#1{}
\begin{document}
\title{Bell States, Dense Coding, and Teleportation}
\author{Subhash Kak\\
Department of Electrical \& Computer Engineering\\
Louisiana State University,
Baton Rouge, LA 70803, USA}
\maketitle

\begin{abstract}
The question of the measurements on the Bell states
by making use of mode change (from mixed to pure) of 
one qubit
is considered. 
Such a mode change cannot be taken advantage of for
superluminal communication in
teleportation, and it may define constraints on the size
of the gates.

\end{abstract}

\thispagestyle{empty}

\subsection*{Introduction}
Underlying protocols for
 dense coding and teleportation\cite{Be95,Pe95} is the 
idea of distinguishing between the four orthogonal Bell states
\vspace{0.2in}

$ |B_1\rangle = \frac{1}{\sqrt 2} (|00\rangle + |11\rangle), $

$ |B_2\rangle = \frac{1}{\sqrt 2} (|00\rangle - |11\rangle), $

$ |B_3\rangle = \frac{1}{\sqrt 2}  (|01\rangle + |10\rangle),$

$ |B_4\rangle = \frac{1}{\sqrt 2}  (|01 \rangle - |10\rangle)$

\noindent
by operations on the entangled qubits.
One may wish to consider these protocols also
to account for effects of mode change for each individual
particle from their original
mixed state
represented
by the density matrix

\vspace{0.2in}
 $                       \left[ \begin{array}{cc}
0.5 & 0  \\
0 & 0.5  \\
\end{array} \right]$
$= \frac{1}{2}(|0\rangle\langle 0| +|1\rangle\langle 1|)$

\vspace{0.2in}
\noindent
to some pure state upon observation of one particle\cite{Ka03}. 
This raises the question if this
transition can be detected gainfully.
This note examines this question.

\subsection*{Dense coding}
In the dense coding protocol\cite{Be95}, the use of an XOR gate
drives the second qubit of $|B_1\rangle$ and $|B_2\rangle$ into $0$
and the second qubit of
$|B_3\rangle$ and $|B_4\rangle$ into $1$.
For example, $|B_1 \rangle$ is transformed into
$\frac{1}{\sqrt 2} (|00\rangle + |10\rangle)= \frac{1}{\sqrt 2}  (|0\rangle + |1\rangle) |0\rangle.$

The measurement of the second qubit 
reduces the state into either $\frac{1}{\sqrt 2} (|0\rangle + |1\rangle)$
or $\frac{1}{\sqrt 2} (|0\rangle - |1\rangle)$.
Since the state may now be factored, each of the two qubits are 
transformed into the pure form.

The transformation 

\vspace{0.2in}
 $H = \frac{1}{\sqrt 2}                       \left[ \begin{array}{cc}
1 & ~1  \\
1 & -1  \\
\end{array} \right]$

\vspace{0.2in}
\noindent
is now applied to this state to distinguish between the terms with
plus and minus.

The capacity of $H$ to distinguish between the two states
(requiring that the state presented to the operator 
 be pure) rests on $\frac{1}{\sqrt 2} (|0\rangle + |1\rangle)$
becoming $|0\rangle$ and
$\frac{1}{\sqrt 2} (|0\rangle - |1\rangle)$ becoming $|1\rangle$. 
This provides the second classical bit of information.

As mentioned before, the application of the XOR changes the individual qubit
state functions from mixed to pure, and
this change of mode may be taken advantage of to communicate
information about the gate operation without collapsing 
the state function.

However this communication will only occur across the size
of the physical gate.
Since no communication can be faster than the speed of light,
and the transition of the state function from mixed to pure
is considered to take place instantaneously, this may 
imply limits on the size of the quantum gate.

\subsection*{Teleportation}
This protocol is the converse of the dense coding protocol.
Here Alice and Bob start with the entangled pair
$ | B_1\rangle =  \frac{1}{\sqrt 2} (| 00 \rangle +  | 11\rangle)$.
Alice wishes to send to Bob the unknown qubit
$|\phi\rangle$.
Without loss of generality, $|\phi\rangle = a |0\rangle + b |1\rangle$, where 
$a$ and $b$ are unknown coefficients. The initial state of the three
qubits is:

\vspace{0.2in}

$ a |000\rangle + b |100\rangle + a |011\rangle + b |111\rangle $

\vspace{0.2in}
\noindent
(For convenience, we leave out the constant factor in this and
other expressions that follow.)

Alice now applies the XOR operator on the
first two qubits,  obtaining the state:

\vspace{0.2in}

$  a |000\rangle + b |110\rangle + a |011\rangle + b |101\rangle $

\vspace{0.2in}
This is now followed by the $H$ operator on the first qubit, giving us:
\vspace{0.2in}

$ a (|000\rangle +  |100\rangle ) + b  (|010\rangle -  |110\rangle) $

$ + a  (|011\rangle +  |111\rangle ) + b ( |001\rangle -  |101\rangle) $

\vspace{0.2in}
\noindent
Simplifying, we obtain:

\vspace{0.2in}

$ |00\rangle ( a |0\rangle + b |1\rangle )
+ |01\rangle (a |1\rangle + b |0\rangle )$
\vspace{0.1in}

$ + |10\rangle (a |0\rangle - b |1\rangle )
+ |11\rangle (a |1\rangle - b |0\rangle )$

\vspace{0.2in}

Alice now measures the first two qubits (the unknown one and the
first of
the two entangled ones). The state of the remaining qubit
collapses to one of the four states:

\vspace{0.2in}
$a |0\rangle \pm b |1\rangle,$ $ a |1\rangle \pm b |0\rangle$.

\vspace{0.2in}
The information of the two qubits is used in the 
protocol to determine which
of the operators 
$I =      \left[ \begin{array}{cc}
              1  &  \\
                & 1 \\
               \end{array} \right]$,
$A =      \left[ \begin{array}{cc}
                & 1  \\
              1 &  \\
               \end{array} \right]$,
$B =      \left[ \begin{array}{cc}
              1  &  \\
                & -1 \\
               \end{array} \right]$,
$C =      \left[ \begin{array}{cc}
                & 1  \\
              -1 &  \\
               \end{array} \right]$
should be applied to his qubit to place it in the
state $|\phi\rangle$.

\vspace{0.2in}

Although the operators XOR and H did not interact
with this third qubit, its state changed because
of the entanglement and
the measurements of the other qubits.

\subsection*{Teleportation of mixed state}

The
state $|\phi\rangle$ need not be pure.
If the original state is mixed then the teleported state
will also be mixed.

Consider Alice can pick photons in states $|0\rangle$ and
$|1\rangle$ randomly and send them to Bob using the
teleportation protocol.

If Alice picks $|0\rangle$, the final state is:

\vspace{0.2in}
$ |00\rangle |0\rangle + |01\rangle |1\rangle + |10\rangle |0\rangle + |11\rangle |1\rangle $

\vspace{0.2in}

\noindent
If she picks $|1\rangle$, the final state is:

\vspace{0.2in}
$ |00\rangle |1\rangle + |01\rangle |0\rangle - |10\rangle |1\rangle - |11\rangle |0\rangle $

\vspace{0.2in}
Although any specific photon will be in a pure state after the measurement,
Bob will represent it as a mixture, indicating
that the teleported mixed state may not necessarily
represent the state of the specific copy of the input.
Should Alice inform Bob later and tell which kind of
photon had been transmitted, the state function of that photon
will change so far as Bob is concerned.  

\subsection*{Mode change in teleportation}

Suppose one begins with the state function $|\phi\rangle
= \frac{1}{\sqrt 2} (|0\rangle + |1\rangle )$.
The mode change occurs in Bob's qubit as soon as
the measurements of the first two qubits are made by
Alice. 
But Bob's qubit, when it switches to the pure mode, will
be in one of the two different orthogonal states 
$\frac{1}{\sqrt 2} (|0\rangle + |1\rangle )$ and
$\frac{1}{\sqrt 2} (|0\rangle - |1\rangle )$,
making it impossible to distinguish these from the
earlier mixed state before measurements by Alice
have been communicated to Bob on the classical channel to
help him distinguish between the two orthogonal states.

\subsection*{Conclusion}
This note takes a look at the dense coding and teleportation protocols
from the point of view of exploiting mode change, when the state function 
changes from mixed to pure, or from one mixed state to another (if the
original
state is mixed in the teleportation protocol).
Expectedly, such a transition cannot be taken advantage of to transmit information 
faster than the speed of light. 
It may define constraints on the physical size of the quantum
gates.


\end{document}